\documentclass[paper]{JHEP3} % 10pt is ignored!

\usepackage{epsfig,multicol}

\newcommand{\pp}{{=\!\!\!|}}
\newcommand\fverb{\setbox\pippobox=\hbox\bgroup\verb}
\newcommand\fverbdo{\egroup\medskip\noindent%
            \fbox{\unhbox\pippobox}\ }
\newcommand\fverbit{\egroup\item[\fbox{\unhbox\pippobox}]}
\newbox\pippobox
\title{Supersymmetric non-linear $ \sigma $-models with boundaries revisited}

\author{Paul Koerber\thanks{Aspirant FWO} , Stijn Nevens${}^*$ and Alexander
Sevrin \\
Theoretische Natuurkunde, Vrije Universiteit Brussel \\
Pleinlaan 2, B-1050 Brussels, Belgium \\
        E-mail: \email{koerber@tena4.vub.ac.be}, \email{stijn@tena4.vub.ac.be},
    \email{asevrin@tena4.vub.ac.be}}

\preprint{\hepth{0309229}}  % OR: \preprint{Aaaa/Mm/Yy\\Aaa-aa/Nnnnnn}
\abstract{We study two-dimensional supersymmetric non-linear $ \sigma $-models with boundaries.
We derive the most general family of boundary conditions in the non-supersymmetric case. Next
we show that no further conditions arise when passing to the $N=1$ model. We present
a manifest $N=1$ off-shell
formulation. The analysis is greatly simplified compared to previous studies and
there is no need to introduce non-local superspaces nor to
go (partially) on-shell. Whether or not torsion is present does not modify
the discussion. Subsequently, we determine under which conditions a second
supersymmetry exists. As for the case without boundaries, two covariantly constant
complex structures are needed. However, because of the presence of the boundary, one
gets expressed in terms of the other one and the remainder of the geometric data.
Finally we recast some of our results in $N=2$ superspace and discuss applications.}


\begin{document}

\section{Introduction}
Non-linear $ \sigma $-models in two dimensions with $N=(2,2)$ supersymmetries play
a central role in the description of strings in non-trivial NS-NS backgrounds.
In the absence of boundaries, a case relevant to type II strings, their geometry has been
intensively studied in the past, see {\em e.g.}
\cite{luis}--\cite{BSVV}. Much less is known for the case with boundaries which is relevant for
type I string theories and D-branes. Partial results were known for some time, see {\em e.g.}
\cite{Ooguri:1996ck}--\cite{Haggi-Mani:2000uc}, however only recently a systematic study
was performed, \cite{stock1}--\cite{stock2}, resulting in the most general boundary conditions compatible
with $N=1$ supersymmetry. Subsequently, these results were extended to $N=2$ supersymmetry \cite{zab}
(see also \cite{Lindstrom:2002vp} for some specific applications and \cite{Lindstrom:2002mc} for a different approach).

While impressive, the results of \cite{stock1} and \cite{stock2} remain somewhat surprising.
Not only are the derivations quite
involved, but the presence of a Kalb-Ramond background seems to require a non-local superspace description
of the model. This already occurs in the very simple setting where open strings move in
a trivial gravitational background but in a non-trivial
electro-magnetic background. It is clear that in order to study the open string
effective action through the calculation of supergraphs, the original motivation for the present
investigation, a local superspace description is called for.

In this paper we reanalyze the models studied in \cite{stock1} and \cite{stock2} and we resolve many of the
difficulties encountered there. We start by reconsidering a non-supersymmetric non-linear $ \sigma $-model and study
the most general boundary conditions. In the next section we extend this to models with supersymmetry.
Motivated by the methods used in \cite{oliver} and \cite{joanna} (in quite a different setting however),
we use a superspace formulation which is manifestly
invariant under only one combination of the two bulk supersymmetries. In this way the analysis of boundary conditions
compatible with $N=1$ supersymmetry is greatly facilitated and one finds that, just as for the case without boundaries,
$N=0$ automatically implies $N=1$. In addition, no non-local terms are needed and the cases
with or without Kalb-Ramond background are treated on the same footing. The price we pay for this is that we loose manifest bulk $d=2$
super Lorentz covariance.

Next we investigate under which conditions the $N=1$ supersymmetry
gets promoted to an $N=2$ supersymmetry. As for the case without
boundaries, one needs two separately integrable covariantly
constant complex structures. The metric has to be hermitian with
respect to both of them. However, the presence of boundaries
requires that one of them gets expressed in terms of the other one
and the remainder of the geometric data.

Finally, we briefly study the $N=2$ superspace formulation.

\section{No supersymmetry}
Varying the bosonic two-dimensional non-linear $ \sigma $-model action\footnote{Derivatives
with respect to $ \tau $ and $ \sigma $ are denoted by a dot and a prime respectively.},
\begin{eqnarray}
{\cal S}=\int d \tau d \sigma \left( \frac{1}{2}\dot X{}^a\, g_{ab}\,\dot X{}^b-
\frac{1}{2} X{}^a{}'\, g_{ab}\, X{}^b{}'+X^a{}'\,b_{ab}\,\dot X^b
\right),\label{bosac}
\end{eqnarray}
we get, apart from the well-known bulk contribution, a boundary term\footnote{When describing open strings, one deals with two boundaries. The
present discussion is readily generalized to this case.},
\begin{eqnarray}
\int d \tau\, \delta X^a \,g_{ab}\,\left(- X^b{}'+b^b{}_c\,\dot X^c\right).\label{bbdy}
\end{eqnarray}
The boundary term vanishes if we either impose Neumann boundary conditions in all directions,
\begin{eqnarray}
X^a{}'-b^a{}_b\,\dot X^b=0,
\end{eqnarray}
or Dirichlet boundary conditions in all directions,
\begin{eqnarray}
\delta X^a=0.
\end{eqnarray}
In order to introduce mixed boundary conditions we need
a $(1,1)$-tensor $ {\cal R}^a{}_b(X)$ satisfying,
\begin{eqnarray}
{\cal R}^a{}_c{\cal R}^c{}_b= \delta ^a{}_b.\label{RRisone}
\end{eqnarray}
This allows us to construct the projection operators ${\cal P}_\pm$,
\begin{eqnarray}
{\cal P}_\pm^a{}_b\equiv \frac{1}{2}\left( \delta ^a{}_b\pm {\cal R}^a{}_b\right).\label{bproj}
\end{eqnarray}
With this we impose simultaneously Neumann,
\begin{eqnarray}
{\cal P}_+^a{}_b \left (X^b{}'-b^b{}_c\,\dot X^c\right)=0,\label{bbn}
\end{eqnarray}
and Dirichlet boundary conditions,
\begin{eqnarray}
{\cal P}_-^a{}_b \delta X^b=0.\label{bbd}
\end{eqnarray}
In other words $ {\cal P}_+$ and $ {\cal P}_-$, project onto
Neumann and Dirichlet directions respectively.
The boundary conditions, eqs.\ (\ref{bbn}) and (\ref{bbd}), can also be rewritten as,
\begin{eqnarray}
X^a{}'&=& {\cal P}_-^a{}_bX^b{}'+ {\cal P}^a_+{}_bb^b{}_c {\cal P}^c_+{}_d\dot X^d, \nonumber\\
\delta X^a&=& {\cal P}_+^a{}_b \delta X^b.\label{rewri}
\end{eqnarray}
It is not hard to see that using these conditions the boundary term eq.\ (\ref{bbdy}) vanishes
provided the metric is invariant under the $(1,1)$ tensor,
\begin{eqnarray}
{\cal R}^c{}_a {\cal R}^d{}_bg_{cd}=g_{ab}.\label{metinv}
\end{eqnarray}
Using eq.\ (\ref{RRisone}) this gives $ {\cal R}_{ab}= {\cal R}_{ba}$.

Requiring time independence, {\em i.e.} if $X^a(\tau,\sigma)$ satisfies the boundary conditions,
so should $X^a(\tau+\delta \tau,\sigma)$, we can put $\delta X^b=\dot{X}^b \delta \tau$ in
eq.\ (\ref{bbd}) and find,
\begin{eqnarray}
{\cal P}_-^a{}_b \dot{X}^b=0.\label{bbd2}
\end{eqnarray}
Using then $[ \delta , \partial/ \partial\, \tau ]=0$ on the boundary,
\begin{eqnarray}
0 = [ \delta , \partial/ \partial\, \tau ] X^c = 2 {\cal P}_+^d{}_{[a} {\cal P}_+^e{}_{b]} {\cal P}^c_+{}_{d,e} \delta X^a \dot{X}^b,\label{deldel}
\end{eqnarray}
yields the condition,
\begin{eqnarray}
{\cal P}_+^d{}_{[a} {\cal P}_+^e{}_{b]} {\cal P}^c_+{}_{d,e}=0.\label{integrability}
\end{eqnarray}
% Using the last of eq.\ (\ref{rewri}), one rewrites the commutator of two infinitesimal
% vectors as,
% \begin{eqnarray}
% {\cal P}_-^a{}_b \left(
% \delta _1X^c \delta _2X^b{}_{,c}-
% \delta _2X^c \delta _1X^b{}_{,c}
% \right)=2\,
% {\cal P}^a_+{}_{[d,e]}{\cal P}_+^d{}_{b} {\cal P}_+^e{}_{c}
% \delta _2X^b \delta _1X^c.\label{deldel}
% \end{eqnarray}
% Requiring that on the boundary, the commutator of two displacements in the Neumann direction
% remains in the Neumann direction, yields the condition\footnote{Note that
% ${\cal P}^c_+{}_{d,e}{\cal P}_+^d{}_{[a} {\cal P}_+^e{}_{b]}=
% {\cal P}_-^c{}_f{\cal P}^f_+{}_{d,e}{\cal P}_+^d{}_{[a} {\cal P}_+^e{}_{b]}$},
% \begin{eqnarray}
% {\cal P}_+^d{}_{[a} {\cal P}_+^e{}_{b]} {\cal P}^c_+{}_{d,e}=0.\label{integrability}
% \end{eqnarray}
%Eq.\ (\ref{integrability}) also guarantees that $[ \delta , \partial/ \partial\, \tau ]=0$ holds both in the bulk and on the boundary.
The necessity of eq.\ (\ref{integrability}) can also be seen in the case where $b$ is exact, $b_{ab}=
\partial_aA_b- \partial_b A_a$. Then using eq.\ (\ref{bbd2}), one can rewrite eq.\ (\ref{bosac}) as,
\begin{eqnarray}
{\cal S}=\int d \tau d \sigma \left( \frac{1}{2}\dot X{}^a\, g_{ab}\,\dot X{}^b-
\frac{1}{2} X{}^a{}'\, g_{ab}\, X{}^b{}'
\right)+\int d \tau A_a {\cal P}_+^a{}_b\dot X^b.\label{bosac1}
\end{eqnarray}
Varying eq.\ (\ref{bosac1}), one indeed obtains eq.\ (\ref{bbn}) provided eq.\ (\ref{integrability}) holds.
As a consequence, the integrability condition, eq. (\ref{integrability}), implies that the commutator 
of two infinitesimal displacements in the Neumann direction remains in the Neumann direction\footnote{Note that
${\cal P}^c_+{}_{d,e}{\cal P}_+^d{}_{[a} {\cal P}_+^e{}_{b]}=
{\cal P}_-^c{}_f{\cal P}^f_+{}_{d,e}{\cal P}_+^d{}_{[a} {\cal P}_+^e{}_{b]}$},
\begin{eqnarray}
{\cal P}_-^a{}_b \left(
\delta _1X^c \delta _2X^b{}_{,c}-
\delta _2X^c \delta _1X^b{}_{,c}
\right)=2\,
{\cal P}^a_+{}_{[d,e]}{\cal P}_+^d{}_{b} {\cal P}_+^e{}_{c}
\delta _2X^b \delta _1X^c=0.\label{comdeldel}
\end{eqnarray}

Summarizing, we can have Neumann, eq.\ (\ref{bbn}), and Dirichlet, eq.\ (\ref{bbd}), boundary conditions,
provided there exists a $(1,1)$-tensor $ {\cal R}$ which satisfies,
\begin{eqnarray}
&&{\cal R}^a{}_c {\cal R}^c{}_b= \delta ^a{}_b, \label{c1} \\
&& {\cal R}^c{}_a {\cal R}^d{}_b g_{cd}=g_{ab}, \label{c2}\\
&& {\cal P}_+^d{}_{[a} {\cal P}_+^e{}_{b]} {\cal P}^c_+{}_{d,e}=0. \label{c3}
\end{eqnarray}
Eq.\ (\ref{c1}) tells us that $ {\cal R}$ is an almost product structure for which the metric is preserved,
eq.\ (\ref{c2}). The last condition, eq.\ (\ref{c3}), tells us that the projection operator $ {\cal P}_+$ is integrable. Note that this is
weaker than requiring that $ {\cal R}$ is integrable. The latter would require that
\begin{eqnarray}
{\cal R}^a{}_d{\cal R}^d{}_{[b,c]}+{\cal R}^d{}_{[b}{\cal R}^a{}_{c],d}=0,
\end{eqnarray}
holds. This is equivalent to the integrability of both $ {\cal P}_+$ and $ {\cal P}_-$ as can be seen from
\begin{eqnarray}
{\cal P}_+^d{}_{[b} {\cal P}_+^e{}_{c]} {\cal P}^a_+{}_{d,e}=
-2 {\cal P}_-^a{}_e\left({\cal R}^e{}_d{\cal R}^d{}_{[b,c]}+{\cal R}^d{}_{[b}{\cal R}^e{}_{c],d}\right).
\end{eqnarray}

\section{$N=1$ supersymmetry}
\subsection{The superspace formulation}
\label{N1superspace}
We work in $N=1$ superspace with a single real fermionic coordinate $ \theta $ and we have
the supersymmetry generator $Q$ and fermionic derivative $D$ such that,
\begin{eqnarray}
Q^2=+\frac i 4 \frac{ \partial\ }{ \partial \tau }\,,\qquad D^2=-\frac i 4 \frac{ \partial\ }{ \partial \tau }\,.\label{qd2}
\end{eqnarray}
We introduce bosonic $N=1$ superfields $X^a$ and fermionic superfields $\Psi^a$, $a\in\{1,\cdots , D\}$.
{From} the point of view of the target
manifold the former will be coordinates while the latter are vectors. The $X$ superfields contain the bulk scalar fields and half of the
bulk fermionic degrees of freedom. The $ \Psi $ superfields contain the other half of the bulk fermionic fields and the auxiliary fields.
In this section, we will stick to Neumann boundary conditions and postpone the analysis of more general boundary conditions to
the next subsection.
On the boundary, the fermionic degrees of freedom are halved and no auxiliary fields are needed anymore. In other words, the boundary
conditions should be such that $ \Psi $ gets expressed as a function of $X$ so that only $X$ lives on the boundary.
In order to do so we assume that the
$\sigma$ derivatives act only on the $X$ superfields. In this way only the variation of $X$ will result in a boundary term.
We take as action,
\begin{eqnarray}
{\cal S}=\int d \tau d \sigma d \theta \sum_{j=1}^{10}{\cal L}_{(j)}, \label{Sd1}
\end{eqnarray}
where,
\begin{eqnarray}
{\cal L}_{(1)}&=& {\cal O}^{(1)}_{ab}(X)DX^a \dot X^b ,\nonumber\\
{\cal L}_{(2)}&=& {\cal O}^{(2)}_{ab}(X) DX^a  X^b{}',\nonumber\\
{\cal L}_{(3)}&=& {\cal O}^{(3)}_{ab}(X) \Psi^a X^b{}' ,\nonumber\\
{\cal L}_{(4)}&=& {\cal O}^{(4)}_{ab}(X) D \Psi ^a \Psi ^b,\nonumber\\
{\cal L}_{(5)}&=& {\cal O}^{(5)}_{ab}(X)D \Psi ^a DX^b ,\nonumber\\
{\cal L}_{(6)}&=& {\cal O}^{(6)}_{ab}(X) \Psi ^a\dot X^b,\nonumber\\
{\cal L}_{(7)}&=& {\cal T}^{(1)}_{abc}(X) \Psi ^a\Psi^b \Psi^c , \nonumber\\
{\cal L}_{(8)}&=& {\cal T}^{(2)}_{abc}(X) \Psi^a \Psi^b DX^c,\nonumber\\
{\cal L}_{(9)}&=& {\cal T}^{(3)}_{abc}(X) \Psi^a DX^b DX^c,\nonumber\\
{\cal L}_{(10)}&=& {\cal T}^{(4)}_{abc}(X)DX^a DX^b DX^c,\label{Ld1}
\end{eqnarray}
where both $ {\cal O}^{(j)}_{ab}(X)$ and $ {\cal T}^{(j)}_{abc}(X)$ are a priori undetermined functions
of $X^a$. On dimensional grounds, it is not hard to see that this is the most general action we can write down
under the assumption that only $X'$ and not $ \Psi '$ appears. Any other term one can write reduces upon
partial integrating $D$ or $ \partial/ \partial \tau $ to the terms listed above. In the next we are going to simplify
eq.\ (\ref{Ld1}) as much as possible.
\begin{itemize}
\item By varying $X$, one immediately gets the boundary condition,
\begin{eqnarray}
{\cal O}^{(3)}_{ba}(X) \Psi^b = -{\cal O}^{(2)}_{ba}(X) DX^b .
\end{eqnarray}
In order that this identifies $ \Psi $ in terms of $X$, we require that ${\cal O}^{(3)}$ is invertible.
\item Performing the integral over $ \theta $, one finds that $D \Psi |_{ \theta =0}$ is auxiliary. Requiring that
the auxiliary field
equations of motion can be solved for the auxiliary fields necessitates that ${\cal O}^{(4)}$ is invertible as well.
\item Partially integrating $D$ in ${\cal L}_{(4)}$ shows that ${\cal O}^{(4)}_{[ab]}$ can be absorbed in $ {\cal T}^{(2)}$.
As such we take from now on ${\cal O}^{(4)}_{ab}={\cal O}^{(4)}_{ba}$.
\item Performing a field redefinition $ \Psi ^a\rightarrow \Psi ^a+{\cal N}^a{}_b(X) DX ^b$ affects almost all terms except
${\cal O}^{(3)}$ and ${\cal O}^{(4)}$. Of particular interest to us is the effect on ${\cal O}^{(5)}$ and ${\cal O}^{(6)}$,
\begin{eqnarray}
{\cal O}^{(5)}_{ab}&\rightarrow& {\cal O}^{(5)}_{ab}+{\cal O}^{(4)}_{ac} {\cal N}^c{}_b, \nonumber\\
{\cal O}^{(6)}_{ab}&\rightarrow& {\cal O}^{(6)}_{ab}-\frac i 4 {\cal N}^c{}_b{\cal O}^{(4)}_{ca} .\label{fr1}
\end{eqnarray}
\item After this, we can completely eliminate $ {\cal L}_{(5)}$ through partial integration which affects
${\cal T}^{(3)}$ and modifies ${\cal O}^{(6)}$ on top of eq.\ (\ref{fr1}) to,
\begin{eqnarray}
{\cal O}^{(6)}_{ab}&\rightarrow& {\cal O}^{(6)}_{ab}-\frac i 4 {\cal O}^{(5)}_{ab}-\frac i 2 {\cal N}^c{}_b{\cal O}^{(4)}_{ca} .\label{fr2}
\end{eqnarray}
\item As we already mentioned, ${\cal O}^{(4)}$ is invertible, so one sees from eq.\ (\ref{fr2}) that a suitable
choice for $ {\cal N}$ can be found such that $ {\cal L}_{(6)}$ vanishes as well.
\item Rewriting $ \partial/ \partial \tau = 4i D^2$ and partially integrating one $D$ shows that $ {\cal O}^{(1)}_{[ab]}$ can
be absorbed in $ {\cal T}^{(4)}$. So from now we take we take ${\cal O}^{(1)}_{ab}={\cal O}^{(1)}_{ba}$.
\item We are still free to perform a field redefinition of the form $ \Psi ^a\rightarrow {\cal M}^a{}_b(X) \Psi ^b$. After this,
$ {\cal O}^{(1)}$ and $ {\cal O}^{(2)}$ are unchanged, $ {\cal O}^{(5)}$ and $ {\cal O}^{(6)}$ remain zero and $ {\cal O}^{(3)}$
and $ {\cal O}^{(4)}$ transform as,
\begin{eqnarray}
{\cal O}^{(3)}_{ab} &\rightarrow& {\cal M}^c{}_a{\cal O}^{(3)}_{cb}, \nonumber\\
{\cal O}^{(4)}_{ab} &\rightarrow& {\cal M}^c{}_a{\cal O}^{(4)}_{cd}{\cal M}^d{}_b.
\end{eqnarray}
As ${\cal O}^{(3)}$ is invertible, we can make a suitable choice for $ {\cal M}$ such that $O^{(3)}_{ab}=-4\, g_{ab}$, with $g_{ab}$ the
target space metric. With this, we exhausted the field redefinitions of $ \Psi $.
\end{itemize}
Concluding we found that without any loss of generality we can put in eq.\ (\ref{Ld1}),
\begin{eqnarray}
{\cal O}^{(1)}_{[ab]}={\cal O}^{(4)}_{[ab]}=0,\qquad {\cal O}^{(5)}_{ab}={\cal O}^{(6)}_{ab}=0,
\qquad {\cal O}^{(3)}_{ab}=-4\,g_{ab}.\label{res1}
\end{eqnarray}
We now proceed with the comparison of eqs.\ (\ref{Sd1}), (\ref{Ld1}) and (\ref{res1}) with the non-linear $ \sigma $-model lagrangian
eq.\ (\ref{lag11}). We assume that the auxiliary fields have already been eliminated in eq.\ (\ref{lag11}). Performing the $ \theta $ integral
in eq.\ (\ref{Sd1}) and eliminating the auxiliary fields $D \Psi ^a|_{ \theta =0}$, one finds by identifying the leading bosonic terms
$\dot X\dot X$, $\dot X X'$ and $X'X'$ to the corresponding terms in eq.\ (\ref{lag11}) that the remaining freedom in the $ {\cal O}$ tensors
gets fully fixed,
\begin{eqnarray}
{\cal O}^{(1)}_{ab}=2i \,g_{ab},\qquad {\cal O}^{(2)}_{ab}=-4i\,b_{ab}, \qquad {\cal O}^{(4)}_{ab}=8 \,g_{ab}.\label{res2}
\end{eqnarray}

Next we want to identify the fermions $DX^a|_{ \theta =0}$ and $ \Psi^a |_{ \theta =0}$ with the bulk fermions $ \psi ^a_+$ and $ \psi ^a_-$
which appear in eq.\ (\ref{lag11}). We first perform the $ \theta $ integral in eq.\ (\ref{Sd1}) using eqs.\ (\ref{Ld1}), (\ref{res1}) and (\ref{res2})
and subsequently eliminate the auxiliary fields. The leading terms quadratic in the fermions in the lagrangian are given by,
\begin{eqnarray}
2i\,g_{ab}\,\Big( i DX^a+ \Psi ^a\Big) \partial_= \Big(i DX^b+ \Psi ^b\Big)+
2i\,g_{ab}\,\Big( i DX^a- \Psi ^a\Big) \partial_\pp \Big(i DX^b- \Psi ^b\Big),
\end{eqnarray}
where we used the boundary condition which follows from the lagrangian,
\begin{eqnarray}
\Psi ^a=i\,b^a{}_bDX^b.
\end{eqnarray}
Comparing this to the corresponding terms in eq.\ (\ref{lag11}), we identify,
\begin{eqnarray}
\psi ^a_+&=& i DX^a+ \Psi ^a, \nonumber\\
\psi^a_-&=& \eta (i DX^a- \Psi ^a) ,\label{id1}
\end{eqnarray}
where $\eta \in\{+1,-1\}$ allows one to differentiate between Ramond and Neveu-Schwarz boundary conditions. It will
not play an essential role in this paper.

Next, we determine the ${\cal T}$ tensors by comparing the rest of the terms to eq.\ (\ref{lag11}).
A somewhat tedious but straightforward calculation\footnote{See appendix C for a shortcut.} yields a unique solution,
\begin{eqnarray}
\label{res3}
{\cal T}^{(1)}_{abc}=-\frac{8i}{3}T_{abc}, \quad {\cal T}^{(2)}_{abc}= 8 \Big\{[ba]c\Big\},
\quad {\cal T}^{(3)}_{abc} = -8i T_{abc}, \quad {\cal T}^{(4)} = 0.
\end{eqnarray}

With this we have the full model in $N=1$ superspace. Its lagrangian is explicitly given by,
\begin{eqnarray}
{\cal L}&=&2ig_{ab}DX^a \dot X^b
-4i b_{ab} DX^a  X^b{}'
-4 g_{ab} \Psi^a X^b{}'
+8g_{ab} \nabla \Psi ^a \Psi ^b \nonumber\\
&&-\frac{8i}{3}T_{abc}\Psi ^a\Psi^b \Psi^c
-8iT_{abc}\Psi^a DX^b DX^c,\label{finac}
\end{eqnarray}
where the covariant derivative $\nabla \Psi^a $ is given by,
\begin{eqnarray}
\nabla \Psi ^a=D \Psi ^a+\Big\{{}^a{}_{bc}\Big\} DX^c \Psi ^b .
\end{eqnarray}

\subsection{The boundary conditions}
Varying eq.\ (\ref{finac}) yields a boundary term,
\begin{eqnarray}
-4\int d \tau d \theta  \left(\Psi^a g_{ab}+iDX^ab_{ab}\right)\delta X^b.
\end{eqnarray}
This vanishes if we take Neumann boundary conditions in all directions,
\begin{eqnarray}
\Psi ^a=i\,b^a{}_bDX^b,
\end{eqnarray}
or Dirichlet boundary conditions in all directions,
\begin{eqnarray}
\delta X^a=0.
\end{eqnarray}
The more general case which involves both Dirichlet and Neumann boundary conditions requires the introduction
of an almost product structure $ {\cal R}^a{}_b(X)$ satisfying eq.\ (\ref{RRisone}). Using the projection
operators defined in eq.\ (\ref{bproj}), we impose Neumann,
\begin{eqnarray}
{\cal P}_+^a{}_b \left (\Psi ^b-i\,b^b{}_cDX^c \right)=0,\label{bn}
\end{eqnarray}
and Dirichlet boundary conditions,
\begin{eqnarray}
{\cal P}_-^a{}_b \delta X^b=0.\label{bd}
\end{eqnarray}
{From} eqs.\ (\ref{bd}) and (\ref{bn}), we obtain,
\begin{eqnarray}
\delta X^a&=& {\cal P}^a_+{}_b \delta X^b, \nonumber\\
\Psi ^a&=& {\cal P}^a_-{}_b \Psi ^b
+i {\cal P}^a_+{}_bb^b{}_c {\cal P}_+^c{}_d DX^d.\label{bfin}
\end{eqnarray}
{From} this one observes that, as was to be expected, $ \delta X$ is completely frozen in the
Dirichlet directions while $ \Psi $ gets a component in the Neumann directions when
there is a non-trivial Kalb-Ramond background.
%In order to simplify things, we start by investigating the case where there is no Kalb-Ramond background, $b_{ab}=0$.
Eq.\ (\ref{bd}) implies,
\begin{eqnarray}
{\cal P}^a_-{}_b D X^b={\cal P}^a_-{}_b \dot X^b=0.\label{bdex}
\end{eqnarray}
This equation requires that certain compatibility conditions are satisfied which follow from
the second expression in eq.\ (\ref{qd2}). Indeed acting with $D$ on the first equation in eq.\ (\ref{bdex}),
we get\footnote{Essentially we observe here that if $DX$ lies in a Dirichlet direction, then so does $D^2X$. Note
that next to $D^2=-i/4 \, \partial/\partial\tau$, also $[D, \delta ]=0$ and as in eq.\ (\ref{deldel}) $[\delta ,\partial/\partial\tau]=0$ lead to the same condition.},
\begin{eqnarray}
0=-\frac i 4 {\cal P}^a_-{}_b \dot X^b+{\cal P}^a_-{}_{d,e} {\cal P}^d_+{}_b {\cal P}^e_+{}_c D X^c D X^b,
\end{eqnarray}
where we used the first equation in eq.\ (\ref{bfin}). This is indeed consistent with the second expression
in eq.\ (\ref{bdex}) provided,
\begin{eqnarray}
{\cal P}_+^d{}_{[a} {\cal P}_+^e{}_{b]} {\cal P}^c_-{}_{d,e}=0,
\end{eqnarray}
or equivalently eq.\ (\ref{integrability}) holds.

With this, we can rewrite the boundary term in the variation as,
\begin{eqnarray}
&&\int d \tau d \theta  \left(\Psi^a+iDX^cb_c{}^a\right) g_{ab}\delta X^b=
\int d \tau d \theta \left( {\cal P}_+^a{}_c +{\cal P}_-^a{}_c \right)
\left(\Psi^c+iDX^db_d{}^c\right)
g_{ab} \delta X^b \nonumber\\
&&\quad =\int d \tau d \theta
\left( {\cal P}_+^a{}_c
\left(\Psi^c+iDX^db_d{}^c\right)g_{ab} \delta X^b +
\left(\Psi^a+iDX^db_d{}^a\right)g_{ab}
{\cal P}_-^b{}_c\delta X^c\right),
\end{eqnarray}
where, in order to make the last step, we had to impose,
\begin{eqnarray}
{\cal R}_{ab}= {\cal R}_{ba},
\end{eqnarray}
which, using eq.\ (\ref{RRisone}) is equivalent to eq.\ (\ref{metinv}). Imposing the
Neumann, eq.\ (\ref{bn}), and the Dirichlet, eq.\ (\ref{bd}), boundary conditions, the boundary
term in the variation of the action indeed vanishes.

So we conclude that {\em any} $N=0$ non-linear $ \sigma $-model with given boundary conditions,
allows for an $N=1$ supersymmetric extension given in
eq.\ (\ref{finac}). The Neumann and Dirichlet boundary conditions, eqs.\ (\ref{bn}) and (\ref{bd}), require the
existence of an almost  product structure $ {\cal R}$ which satisfies eqs.\ (\ref{c1}-\ref{c3})

We now briefly compare our results to those obtained in \cite{stock1} and \cite{stock2}. In the present derivation,
whether or not a Kalb-Ramond background is present, does not play any role.
When the Kalb-Ramond background vanishes, $b_{ab}=0$, eqs.\ (\ref{c1}-\ref{c3}) precisely agree with
the conditions derived in \cite{stock1}.
However as supersymmetry is kept manifest, the derivation of these conditions are tremendously simplified. Contrary to
\cite{stock1}, we remained off-shell all the time.
A drawback compared to \cite{stock1}, is the loss of manifest $d=2$ bulk super Lorentz covariance in the present
formulation.
For a non-trivial Kalb-Ramond background, the comparison with the results
in \cite{stock2} is a bit more involved.
A first bonus compared to \cite{stock2} is that we here have a regular superspace formulation,
{\em i.e.} non-local superspace terms are not needed here.
Combining eqs.\ (\ref{id1}), (\ref{bdex}) and (\ref{bn}),
we schematically obtain the following boundary condition for the fermions,
\begin{eqnarray}
\psi_-=\eta \, \frac{ {\cal R}-  b_{++}}{1+ b_{++}}\psi_+, \label{bcfermions}
\end{eqnarray}
where $b_{++}{}^a{}_b$ stands for $ {\cal P}_+^a{}_cb^c{}_d {\cal P}_+^d{}_b$. The (1,1)-tensors,
$ {\cal R}$ and $(1+  b_{++})^{-1}( {\cal R}-  b_{++})$, should be identified with the (1,1)-tensors $r$ and
$R$ in \cite{stock2}. It is then straightforward to show that eqs.\ (\ref{c1}-\ref{c3}) imply the conditions in
eq.\ (3.22) of \cite{stock2}.

\section{More supersymmetry}
\subsection{Promoting the $N=1$ to an $N=2$ supersymmetry}
\label{N2susy}
The action, eq.\ (\ref{finac}), is manifestly invariant under the supersymmetry transformation,
\begin{eqnarray}
\delta X^a= \varepsilon Q X^a,\qquad \delta \Psi ^a = \varepsilon Q \Psi ^a,
\end{eqnarray}
where the supersymmetry generator $Q$ was defined in eq.\ (\ref{qd2}).
We now derive the conditions under which the action eq.\ (\ref{finac}) exhibits a second supersymmetry.
The most general transformation rules consistent with dimensions and statistics that we can write down are,
\begin{eqnarray}
\delta X^a &=&\hat\varepsilon {\cal J}^{\ a}_{(1)b}(X)DX^b+ \hat\varepsilon{\cal J}^{\ a}_{(2)b}(X) \Psi  ^b, \nonumber\\
\delta \Psi ^a &=& \hat\varepsilon{\cal K}^{\ a}_{(1)b}(X)D \Psi ^b+ \hat\varepsilon{\cal K}^{\ a}_{(2)b}(X) \dot X ^b+
 \nonumber \\
&& \hat\varepsilon{\cal L}_{(1)bc}^{\ a}(X) \Psi ^b \Psi ^c + \hat\varepsilon{\cal L}_{(2)bc}^{\ a}(X) \Psi ^b DX^c
+ \hat\varepsilon{\cal L}_{(3)bc}^{\ a}(X) DX ^b DX^c .\label{2ndsusy}
\end{eqnarray}
The only other term which could have been added in the variation of $ \Psi $ is $ \hat\varepsilon{\cal K}^{\ a}_{(3)b}(X) X^{'b}$.
We did not add it as it would require us to go on shell when evaluating eq.\ (\ref{2ndsusy}) on the boundary\footnote{Upon redefining
$ {\cal K}_{(1)}$, ${\cal L}_{(1)}$, $ {\cal L}_{(2)}$ and $ {\cal L}_{(3)}$, it is proportional to the $ \Psi $ equation of motion.
We refer the reader to the discussion in appendix C.}.

Requiring the bulk terms in the variation of eq.\ (\ref{finac}) to vanish under eq.\ (\ref{2ndsusy}) yields,
\begin{eqnarray}
{\cal J}_{(1)} & = & \frac{1}{2}(J +\bar{J}), \quad i {\cal J}_{(2)}=-4 {\cal K}_{(2)}= \frac{1}{2}(J-\bar{J}) \, ,\quad
{\cal K}_{(1)}=-\frac{1}{2}(J+\bar{J}),  \nonumber \\
{\cal L}^{a}_{(1)bc}&=&  \frac{i}{2} \left(\partial_{[b} J^a{}_{c]}-\partial_{[b} \bar{J}^a{}_{c]}\right)
- \frac{i}{2}(J+\bar J)^{da}T_{dbc} \, ,  \nonumber \\
{\cal L}^{a}_{(2)bc}&=&- \frac{1}{2} \left(\partial_{b} J^a{}_{c}+\partial_{b} \bar{J}^a{}_{c}+
2(J+\bar J)^{da}\{dbc\}
 \right) \, ,
\nonumber\\
{\cal L}^{a}_{(3)bc} &=& -\frac{i}{2}(J+\bar J)^{da}T_{dbc} \, ,
\label{2ndsusy2}
\end{eqnarray}
while $J$ and $\bar{J}$ satisfy,
\begin{eqnarray}
g_{a(b} \; J^a{}_{c)}& = & g_{a(b} \; \bar{J}^a{}_{c)}=0, \nonumber \\
\nabla^+_c J^a{}_b & = & \nabla^-_c \bar{J}^a{}_b=0 \, .\label{cs1}
\end{eqnarray}
Before investigating the vanishing of the boundary terms in the variation, we impose the supersymmetry algebra.
In particular the first supersymmetry
has to commute with the second one, which is trivially realized because of $\{Q,D\}=0$.
Subsequently, we need that
\begin{eqnarray}
{[} \delta (\hat\varepsilon_1), \delta (\hat\varepsilon_2){]}X^a=\frac i 2 \hat\varepsilon_1\hat\varepsilon_2\dot X^a,\qquad
{[} \delta (\hat\varepsilon_1), \delta (\hat\varepsilon_2){]} \Psi ^a=\frac i 2 \hat\varepsilon_1\hat\varepsilon_2\dot \Psi ^a,
\label{closure}
\end{eqnarray}
holds {\em on-shell}. This is indeed true provided
\begin{eqnarray}
J^a{}_b J^b{}_c = \bar{J}^a{}_b \bar{J}^b{}_c = - \delta^a_c\, , \quad N^a{}_{bc}[J,J]=N^a{}_{bc}[\bar{J},\bar{J}]=0 \, ,\label{cs2}
\end{eqnarray}
with the Nijenhuistensor N[A,B] given by,
\begin{eqnarray}
N^a{}_{bc}[A,B]=A^d{}_{[b}B^a{}_{c],d} + A^a{}_{d} B^d{}_{[b,c]} + B^d{}_{[b}A^a{}_{c],d} +B^a{}_{d} A^d{}_{[b,c]} \, .
\end{eqnarray}
A noteworthy fact is that the algebra closes {\em off-shell}. This has to be contrasted with the case
without boundaries where the $N=(2,2)$ algebra closes off-shell modulo terms proportional to
${[}J, \bar J{]}$ times equations of motion. If we would modify the transformation rules with an equation of motion term as
in eq.\ (\ref{eomterm}), then one indeed obtains off-shell closure modulo terms proportional to ${[}J,\bar J{]}$.

We now turn to the boundary term in the supersymmetry variation of the action. Using an obvious matrix like notation, one shows that
it vanishes provided
\footnote{The integrable projection operator ${\cal P}_+$ defines a {\em foliation}, {\em i.e.} a set of branes
which together fill the whole target space.  We could restrict to one (or two) of these branes and
call its (their total) worldvolume $\gamma$. If we require the endpoints of the open string to lie on
the submanifold $\gamma$, the boundary will always be a part of $\gamma$.
Conditions (\ref{ibdy}) and (\ref{compa}) then only hold on $\gamma$. We will not follow this approach here
and require these conditions on the whole of target space.},
\begin{eqnarray}
&&{\cal P}_-(J-\bar J) {\cal P}_-=0, \nonumber\\
&&{\cal P}_+(J-\bar J) {\cal P}_+= {\cal P}_+{[}b, J + \bar J{]} {\cal P}_++ {\cal P}_+b\,(J-\bar J)b\, {\cal P}_+,\label{ibdy}
\end{eqnarray}
holds.
Invariance of the boundary conditions, eqs.\ (\ref{bn}) and (\ref{bd}), under the $N=2$ supersymmetry transformations requires,
\begin{eqnarray}
&&{\cal P}_-(J+\bar J) {\cal P}_+=- {\cal P}_-(J-\bar J)b {\cal P}_+, \nonumber\\
&& {\cal P}_+ (J+\bar J) {\cal P}_-= {\cal P}_+b(J-\bar J) {\cal P}_-.\label{compa}
\end{eqnarray}
Using the antisymmetry of $J$ and $b$ and the symmetry of $ {\cal R}$, it is clear that the second equation in
eq.\ (\ref{compa}) is the transposed of the first one.
It is surprising that conditions (\ref{ibdy}) and (\ref{compa}) are strictly algebraic.  Indeed, all derivative terms
disappear using the integrability condition eq.\ (\ref{integrability}).
Using these conditions, together with the previously obtained equations, we can express $\bar J$ in terms of $J$,
\begin{eqnarray}
\bar J &=&(1+b_{++})^{-1}(1-b_{++})J_{++}(1+b_{++})(1-b_{++})^{-1}+J_{--}- \nonumber\\
&&(1+b_{++})^{-1}(1-b_{++})J_{+-}-
J_{-+}(1+b_{++})(1-b_{++})^{-1} \nonumber\\
&=& {\cal M}J {\cal M}^{-1},\label{btype}
%&=&\left(
%{\cal P}_+ \frac{1-b_{++}}{1+b_{++}}- {\cal P}_-
%\right)\,J\,
%\left(
%\frac{1+b_{++}}{1-b_{++}} {\cal P}_+- {\cal P}_-
%\right).
\end{eqnarray}
with
\begin{eqnarray}
{\cal M}= \frac{ {\cal R}-b_{++}}{1+b_{++}},\qquad
{\cal M}^{-1}= \frac{ {\cal R}+b_{++}}{1-b_{++}}.\label{defvanm}
\end{eqnarray}
Note that eqs.\ (\ref{cs1}) and (\ref{cs2}) are invariant under $J\rightarrow J$ and $\bar J\rightarrow-\bar J$, while this
change in eqs.\ (\ref{ibdy}) and (\ref{compa}) turns eq.\ (\ref{btype}) into
\begin{eqnarray}
\bar J &=&-(1+b_{++})^{-1}(1-b_{++})J_{++}(1+b_{++})(1-b_{++})^{-1}-J_{--}+ \nonumber\\
&&(1+b_{++})^{-1}(1-b_{++})J_{+-}+
J_{-+}(1+b_{++})(1-b_{++})^{-1} \nonumber\\
&=&- {\cal M}J {\cal M}^{-1}.\label{atype}
\end{eqnarray}
The latter, eq.\ (\ref{atype}), are called A-type boundary
conditions while the former, eq.\ (\ref{btype}) are B-type
boundary conditions. In the above, we used the notation,
$b_{++}\equiv {\cal P}_+b {\cal P}_+$, $J_{-+}\equiv {\cal P}_-J
{\cal P_+}$, etc. Using the conditions involving $J$, it is quite
trivial to show that $\bar J$ is indeed an almost complex
structure under which the metric $g$ is hermitian. However, the
covariant constancy and integrability of $J$ does not imply that
$\bar J$ as given in eq.\ (\ref{btype}) or eq.\ (\ref{atype}) is
covariantly constant or integrable. This imposes further
conditions on the allowed boundary conditions, geometry and
torsion !

In the case there is no torsion, $T=0$, we find from eqs.\ (\ref{cs1}) and (\ref{cs2}) that the
geometry is K\"ahler.  Without further conditions on the geometry, there can be only one independent
K\"ahler form so that $J=\bar{J}$ with either B-type or A-type boundary conditions.
For the B-type boundary conditions, eq.\ (\ref{btype}), one
finds the well known results,
\begin{eqnarray}
[J, {\cal R}]=[J, b_{++}]=0.
\end{eqnarray}
This implies that the D-brane worldvolume is a K\"ahler submanifold
with K\"ahler form $J_{++}$ and that $b$ is a (1,1)-form with respect to $J_{++}$.
Turning to A-type boundary conditions, eq.\ (\ref{atype}), one gets,
\begin{eqnarray}
b_{++}J_{++}b_{++}=J_{++},\qquad J_{--}=b_{++}J_{+-}=J_{-+}b_{++}=0.
\end{eqnarray}
The latter implies the existence of a second almost complex structure $\tilde J$,
\begin{eqnarray}
\tilde J\equiv b_{++}J_{++}+J_{+-}+J_{-+},
\end{eqnarray}
which is integrable in the case of a space filling brane.
The following relation exists between the dimension of the brane and the rank of $b$,
\begin{equation}
{\rm dim}({\rm brane}) = \frac{1}{2} (D + {\rm rank}(b)) \, .
\end{equation}
In the special case $b=0$, also $J_{++}=0$ and the brane worldvolume becomes a lagrangian
submanifold.  For a more detailed treatment we refer to \cite{Becker:1995kb},
\cite{Bershadsky:1995qy} and \cite{zab}.

Concluding, we find that a second supersymmetry is allowed provided two almost complex structures, $J$ and $\bar J$, exist which are
separately integrable and covariantly constant, albeit with two different connections. Till this point, this is exactly equal to the
situation without boundaries. However
when boundaries are present, it turns out that one of the two complex structures can be expressed in the other one and the remainder
of the geometric data.

\subsection{Generalized boundary conditions}

Having at our disposal $J$ and $\bar{J}$,
we can generalize eq.\ (\ref{id1}) to,
\begin{eqnarray}
\psi ^a_+&=& \left(e^{\alpha J}\right)^a{}_b (i DX^b+ \Psi ^b), \nonumber\\
\psi^a_-&=& \eta \,\left( e^{\pm \alpha \bar{J}}\right)^a{}_b (i DX^b- \Psi ^b) ,\label{idnew}
\end{eqnarray}
where $\alpha$ is an arbitrary angle. This amounts to applying an R-rotation to the
original $\psi_+$ and $\psi_-$. Using eqs.\ (\ref{cs1}) and (\ref{cs2}), one can show that
both possibilities are symmetries of the bulk action. However, only one of them survives on the boundary.

For A-type boundary conditions eq.\ (\ref{idnew}) with the minus sign  leaves the boundary action invariant while taking the plus 
sign leads to a new model, with the boundary condition, (\ref{bcfermions}),
replaced by,
\begin{eqnarray}
\psi_- & = & \eta \, e^{ - \alpha \bar{J}} \, \frac{ {\cal R}-  b_{++}}{1+ b_{++}} \, e^{ \alpha J} \psi_+ \nonumber \\
       & = &\eta \, \frac{ {\cal R}-  b_{++}}{1+ b_{++}} \, e^{  2 \alpha J} \psi_+,
\end{eqnarray}
where we used eq.\ (\ref{atype}). For the B-type boundary conditions it's the other way around, eq.\ (\ref{idnew})  with the plus 
sign leaves the boundary action invariant while taking the minus sign leads to a new model, with the boundary condition, (\ref{bcfermions}),
replaced by,
\begin{eqnarray}
\psi_- & = & \eta \, e^{ \alpha \bar{J}} \, \frac{ {\cal R}-  b_{++}}{1+ b_{++}} \, e^{ \alpha J} \psi_+ \nonumber \\
       & = &\eta \, \frac{ {\cal R}-  b_{++}}{1+ b_{++}} \, e^{ 2 \alpha J} \psi_+,
\end{eqnarray}
where we used eq.\ (\ref{btype}).

\subsection{$N=2$ superspace}
The fact that the supersymmetry algebra, eqs. (\ref{2ndsusy}) and (\ref{2ndsusy2}), closes off-shell, hints towards
the existence of an $N=2$ superspace formulation without the need of introducing further auxiliary fields. However,
the structure of eqs. (\ref{2ndsusy}) and (\ref{2ndsusy2}) shows that the constraints on the $N=2$ superfields will be generically
non-linear. As we will limit ourselves to linear constraints, we will need to ``improve'' the transformation rules with terms which vanish on-shell
as in eqs. (\ref{2ndsusya}) and (\ref{2ndsusyb}). However, as is clear from the discussion in appendix C, 
off-shell closure is then only achieved when $J$ and
$\bar J$ commute, thereby imposing important restrictions on the geometry.

We denote the fermionic coordinates of $N=2$ superspace by $ \theta $ and $\bar \theta $. We introduce the fermionic
derivatives $D$ and $\bar D$ which satisfy,
\begin{eqnarray}
\{D,\bar D\}=-i \partial_\tau,\quad D^2=\bar D^2=0.
\end{eqnarray}
We now want to introduce superfields, which upon integrating out the extra fermionic coordinate, reduces to the fields
introduced in section 3.1. We will restrict ourselves to the simplest case where only linear constraints are used.
In order to achieve this we introduce the $N=1$ derivative $\hat D$ (it corresponds to the $D$ in the previous sections)
and the ``extra'' derivative $\check D$,
\begin{eqnarray}
\hat D \equiv \frac 1 2 \left(D+\bar D\right),\qquad \check D=\frac i 2 \left(D-\bar D\right),
\end{eqnarray}
which satisfy,
\begin{eqnarray}
\hat D^2=\check D^2=-\frac i 4 \partial_\tau,\qquad \{\hat D,\check D\}=0.\label{nis2d}
\end{eqnarray}
We introduce the $N=2$ superfields $X^a$ and $ \Psi ^a$. When passing from $N=2$ to $N=1$ superspace, we do not
want to introduce extra auxiliary degrees of freedom. In order to achieve this, the $\check D$-derivatives of the fields
should satisfy constraints. The most general {\em linear} constraints one can write down are,
\begin{eqnarray}
\check D X^a&=& {\cal C}_1^a{}_b\hat D X^b+ {\cal C}_2^a{}_b \Psi ^b, \nonumber\\
\check D \Psi ^a&=& {\cal C}_3^a{}_b\hat D \Psi ^b+ {\cal C}_4^a{}_b \dot X ^b+ {\cal C}_5^a{}_b X^b{}',\label{suspacon}
\end{eqnarray}
where $ {\cal C}_j$, $j\in\{1,\cdots ,5\}$ are constant. Eq.\ (\ref{nis2d}) implies integrability conditions,
\begin{eqnarray}
&&{\cal C}_1^2=-{\bf 1}+4i {\cal C}_2 {\cal C}_4,\quad {\cal C}_3^2=-{\bf 1}+4i {\cal C}_4 {\cal C}_2, \nonumber\\
&& {\cal C}_2 {\cal C}_5= {\cal C}_5 {\cal C}_2=0, \nonumber\\
&& {\cal C}_1 {\cal C}_2= {\cal C}_2 {\cal C}_3,\quad
{\cal C}_3 {\cal C}_5= {\cal C}_5 {\cal C}_1,\quad
{\cal C}_3 {\cal C}_4= {\cal C}_4 {\cal C}_1.\label{cint}
\end{eqnarray}
These integrability conditions allow one to solve the constraints, eq.\ (\ref{suspacon}), in terms of an
unconstrained, fermionic, dimension -1/2 superfield $\Lambda$, and an unconstrained, bosonic, dimension 0
superfield $Y$,
\begin{eqnarray}
X&=&(\check D- {\cal C}_1\hat D)\Lambda+ {\cal C}_2Y, \nonumber\\
\Psi &=&(\check D- {\cal C}_3\hat D)Y+ {\cal C}_4\dot\Lambda+ {\cal C}_5\Lambda'.
\end{eqnarray}
Motivated by the results in appendix C, we propose the following
parameterization for the tensors $ {\cal C}_j$, $j\in\{1,\cdots
5\}$,
\begin{eqnarray}
&& {\cal C}_1=\frac 1 2 (J+\bar J),\qquad {\cal C}_2=-\frac i 2 (J-\bar J),\qquad {\cal C}_3=\frac 1 2 (J+\bar J+K), \nonumber\\
&& {\cal C}_4=-\frac 1 8 (J-\bar J),\qquad {\cal C}_5=-\frac 1 8 (2J+2\bar J+K),
\label{suspacon2}
\end{eqnarray}
where $J^2=\bar J^2=-1$.
In order that eq.\ (\ref{cint}) is satisfied, one needs
\begin{eqnarray}
K^2=-\{J+\bar J,K\},\qquad 2[J,\bar J]=K(J-\bar J)=(\bar J-J)K.
\end{eqnarray}
This has two obvious solutions:
\begin{eqnarray}
K= -2(J+\bar J)\label{sol1},
\end{eqnarray}
or
\begin{eqnarray}
K=0 \mbox{ and } [J,\bar J]=0.\label{sol2}
\end{eqnarray}
%Obviously, eq.\ (\ref{suspacon}) with eq.\ (\ref{suspacon2}) is,
%modulo non-linear terms, identified with the supersymmetry
%transformation rules eq.\ (\ref{2ndsusya}). In other words eq.\
%(\ref{suspacon}) now reads,
%\begin{eqnarray}
%\check D \Phi^a = \check Q \Phi^a, \qquad \Phi^a=(X^a,\Psi^a),
%\label{suspacon3}
%\end{eqnarray}
%where $\check Q$ generates the linear part of eq.\ (\ref{2ndsusya}). It is easy to see that the integrability
%conditions (\ref{cint}) are equivalent to the {\em off-shell} closure conditions (\ref{closure})
%and thus satisfied by the above.  Also, conditions (\ref{suspacon3}) transform covariantly under
%the second supersymmetry (\ref{2ndsusya}) so that they are valid constraints when passing
%from $N=2$ to $N=1$ superspace.
%
%If we stick to linear constraints we read from eqs.\
%(\ref{2ndsusya}) and (\ref{2ndsusyb}) that we need to opt for eq.\
%(\ref{sol2})! In that case, the two commuting integrable
%structures $J$ and $\bar{J}$ are simultaneously diagonalizable. 

Taking the first possibility, eq.~(\ref{sol1}), in eq.~(\ref{suspacon}), one sees, upon
passing to $N=1$ superspace, that this would correspond to a linearized version of the
supersymmetry transformations in eqs.~(\ref{2ndsusy}) and (\ref{2ndsusy2}). Hence, this would only cover
the trivial case of a flat target space.

So if we want to stick to linear constraints we read from eqs.\
(\ref{2ndsusya}) and (\ref{2ndsusyb}) that we need to opt for eq.\
(\ref{sol2})! In that case, the two commuting integrable
structures $J$ and $\bar{J}$ are simultaneously diagonalizable. 
We
choose complex coordinates so that,
\begin{eqnarray}
J^{\alpha}{}_{\beta} & = & \bar{J}^{\alpha}{}_{\beta} = i \delta^{\alpha}{}_{\beta},  \quad
J^{\bar{\alpha}}{}_{\bar{\beta}} = \bar{J}^{\bar{\alpha}}{}_{\bar{\beta}} = - i \delta^{\bar{\alpha}}{}_{\bar{\beta}}, \quad
\quad \alpha,\beta \in\{1,\cdots m\}, \nonumber \\
J^{\mu}{}_{\nu} & = & - \bar{J}^{\mu}{}_{\nu} = i \delta^{\mu}{}_{\nu},  \quad
J^{\bar{\mu}}{}_{\bar{\nu}} = - \bar{J}^{\bar{\mu}}{}_{\bar{\nu}} = - i \delta^{\bar{\mu}}{}_{\bar{\nu}}, \quad
\quad \mu,\nu \in\{1,\cdots n\},
\end{eqnarray}
and all other components vanishing.
In these coordinates, where we denote the bosonic superfield now by $Z$, eq.\ (\ref{suspacon})
with eq.\ (\ref{suspacon2}) takes the form,
\begin{eqnarray}
&&\check D Z^ \alpha =+i\,\hat D Z^ \alpha ,\quad \check DZ^{\bar \alpha }=-i\,\hat D Z^{\bar \alpha },
\nonumber\\
&&\check D \Psi^ \alpha =+i\, \hat D\Psi^ \alpha -\frac i 2 Z^ \alpha {}',\quad
\check D \Psi^{\bar \alpha }=-i\,\hat D\Psi^{\bar \alpha }+\frac i 2 Z^{\bar \alpha }{}',
\quad \alpha \in\{1,\cdots m\},
\label{csf}
\end{eqnarray}
or equivalently,
\begin{eqnarray}
\bar D Z^ \alpha =D Z^{\bar \alpha }=0,\quad \bar D \Psi^ \alpha=\frac 1 2 Z^ \alpha {}',\quad
D \Psi^{\bar \alpha }= \frac 1 2 Z^{\bar \alpha }{}', \label{chiralsf}
\end{eqnarray}
and
\begin{eqnarray}
\check D Z^ \mu =+\Psi ^ \mu ,\quad \check DZ^{\bar \mu }=-\Psi^{\bar \mu },\quad
\check D\Psi^\mu=-\frac i 4 \dot Z^ \mu ,\quad \check D\Psi^{\bar\mu}=+\frac i 4 \dot Z^{\bar \mu },\quad
\mu \in\{1,\cdots n\}.
\label{tcsf}
\end{eqnarray}
Eqs.\ (\ref{csf}) and (\ref{tcsf}) are the boundary analogs of the two-dimensional chiral and twisted
chiral superfields respectively.

We will only consider the case where only one type of superfields is present. Contrary to the case without
boundaries, this yields two different cases. Having only chiral (twisted chiral) superfields results in a
K\"ahler geometry with B(A)-type supersymmetry. Taking exclusively chiral superfields ($n=0$), we introduce
two potentials $K(Z,\bar Z)$ and $V(Z, \bar Z)$ and the action,
\begin{eqnarray}
\int d^2 \sigma d^2 \theta\, K(Z,\bar Z)_{ ,\alpha \bar \beta }\left(-2i D Z^ \alpha \bar D Z^ {\bar \beta }-8i
\Psi^ \alpha \Psi^{\bar \beta }\right)+
\int d \tau d^2 \theta\, V(Z,\bar Z).
\end{eqnarray}
Passing to $N=1$ superspace one gets the action eq.\ (\ref{finac}) with,
\begin{eqnarray}
&&g_{ \alpha \bar \beta }=K_{ ,\alpha \bar \beta },\quad b_{ \alpha \bar \beta }=-\frac 1 2 V_{, \alpha \bar \beta }, \nonumber\\
&&g_{ \alpha  \beta }=g_{\bar \alpha \bar \beta }=0,\quad b_{ \alpha \beta }= b_{\bar \alpha \bar \beta }=0.
\end{eqnarray}
Solving the constraints in terms of unconstrained superfields $ \Lambda $ and $Y$,
\begin{eqnarray}
Z^ \alpha =\bar D \Lambda ^ \alpha ,\quad Z^{\bar \alpha }= D \Lambda ^{\bar \alpha },\quad
\Psi ^ \alpha = \bar D Y^ \alpha + \frac 1 2 \Lambda ^ \alpha {}',\quad
\Psi^{\bar \alpha }=DY^{\bar \alpha }+ \frac 1 2 \Lambda ^{\bar \alpha }{}',
\end{eqnarray}
and varying the action with respect to the unconstrained superfields, we get the boundary term,
\begin{eqnarray}
\int d \tau  d^2 \theta \, \left(\delta \Lambda ^ \alpha \left(-4i K_{, \alpha \bar \beta }\Psi^{\bar \beta }+
V_{, \alpha \bar \beta }\bar DZ^{\bar \beta }\right)
+\delta \Lambda ^{\bar \alpha} \left(4i K_{, \bar\alpha  \beta }\Psi^{ \beta }+
V_{, \bar\alpha \beta }D Z^{\beta }\right)
\right).
\end{eqnarray}
The boundary term vanishes provided we introduce an almost product structure $ {\cal R}$ which satisfies,
\begin{eqnarray}
&&{\cal R}^ \alpha {}_{ \bar \beta } = {\cal R}^{\bar \alpha }{}_ \beta =0, \nonumber\\
&& {\cal R}_{ \alpha \bar \beta }\equiv g_{ \alpha \bar \gamma }{\cal R}^{\bar \gamma }{}_{\bar \beta } =
{\cal R}_{\bar \beta \alpha  }\equiv g_{\bar \beta  \gamma } {\cal R}^ \gamma {}_ \alpha .
\end{eqnarray}
Using the almost product structure to construct projection operators $ {\cal P}_+$ and $ {\cal P}_-$, we find that
the boundary term in the variation indeed vanishes if we impose,
\begin{eqnarray}
&&{\cal P}_-^{ \alpha }{}_{ \gamma } \delta \Lambda ^{ \gamma }=
{\cal P}_-^{ \bar\alpha }{}_{\bar \gamma } \delta \Lambda ^{\bar \gamma }=0, \nonumber\\
&& {\cal P}_+^{ \alpha }{}_{ \beta }\left( \Psi^{ \beta }-\frac i 4 g^{ \beta \bar \gamma }V_{, \bar \gamma \delta }\bar D Z^{ \delta }\right) =
{\cal P}_+^{\bar \alpha }{}_{ \bar\beta }\left( \Psi^{\bar \beta }+\frac i 4 g^{\bar \beta  \gamma }V_{, \gamma \bar\delta }\bar D Z^{\bar \delta }
\right)=0.
\end{eqnarray}
Demanding compatibility of the first two equations with $  {\cal P}_-^ \alpha {}_{ \beta } \delta Z^ \beta  = 
{\cal P}_-^ {\bar \alpha} {}_{ \bar \beta }\delta Z^{\bar \beta}  =0$   requires,
\begin{eqnarray}
{\cal P}^ \alpha _+{}_{ \delta , \bar \varepsilon } {\cal P}^ \delta _+{}_ \beta 
{\cal P}_+^{\bar \varepsilon }{}_{\bar \gamma }={\cal P}^{\bar \alpha} _+{}_{ \bar \delta , \varepsilon } 
{\cal P}^{\bar \delta} _+{}_ {\bar \beta} {\cal P}_+^{ \varepsilon }{}_{ \gamma }=0.
\end{eqnarray}
 Finally from $D Z^ \alpha = {\cal P}^ \alpha _+{}_ \beta D Z^ \beta $ and $D^2=0$ and likewise from 
 $\bar D \bar Z^ {\bar \alpha} = {\cal P}^{\bar \alpha} _+{}_ {\bar \beta} \bar D \bar Z^ {\bar \beta} $ and $\bar D^2=0$, we get,
\begin{eqnarray}
{\cal P}^ \alpha _+{}_{ [\delta , \varepsilon ]} {\cal P}^ \delta _+{}_ \beta 
{\cal P}_+^{ \varepsilon }{}_{ \gamma }={\cal P}^{\bar \alpha} _+{}_{ [\bar \delta , \bar \varepsilon] } 
{\cal P}^{\bar \delta} _+{}_ {\bar \beta} {\cal P}_+^{ \bar \varepsilon }{}_{\bar \gamma }=0.
\end{eqnarray}
The conditions obtained here are completely equivalent to those in eqs.\ (\ref{c1}-\ref{c3}) and (\ref{btype}) for a K\"ahler geometry.

We now briefly turn to the case where we take exclusively twisted chiral superfields ($m=0$). The action,
\begin{eqnarray}
{\cal S}=\int d^2 \sigma d^2 \theta \left(-8K_{,\mu\bar\nu}\Psi^{\mu}\hat D Z^{\bar\nu}+
8K_{,\bar\mu\nu}\Psi^{\bar\mu}\hat D Z^{\nu}
+ 2 K_{,\mu} Z^{\mu'} - 2 K_{,\bar{\mu}} Z^{\bar{\mu}'} \right),
\end{eqnarray}
correctly reproduces the bulk theory, however it does not give the right boundary terms. In other words, the
$N=2$ superspace description of type A boundary conditions remains unknown.
%is equivalent to the one in eq.\ (\ref{finac}) with $b=0$ and $g_{\mu\nu}=g_{\bar\mu\bar\nu}=0$, $g_{\mu\bar\nu}=K_{,\mu\bar\nu}$. 
%Unfortunately, despite the fact that a more detailed analysis shows that this does describe type A boundary conditions, 
%this case turns out to be not particularly interesting as non-trivial $U(1)$ backgrounds cannot be described using these superfields.

\section{Conclusions/discussion}
In this paper we studied $d=2$ non-linear $ \sigma $- models in the presence of boundaries. In the absence of supersymmetry,
we found that the boundary conditions require the existence of an almost product structure $ {\cal R}$ compatible with the metric and such that
the projector $ {\cal P}_+=(1+ {\cal R})/2$ is integrable. Supersymmetrizing the model yields no further conditions. We obtained
a manifest $N=1$ supersymmetric formulation of the model. Whether or not torsion is present does not essentially alter the discussion.

Crucial in this was $ {\cal M}$, eq.~(\ref{bcfermions}) (see also eq.~(\ref{defvanm})) which relates the left movers to the right
movers. In the case of constant magnetic background fields it was for the first time written down in \cite{Callan:1988wz}. In 
terms of formal power series we can rewrite it in the following form,
\begin{eqnarray}
{\cal M}&=& \frac{1-b_{++}}{1+b_{++}} {\cal P}_+- {\cal P}_- \nonumber\\
&=&e^{-2\,\mbox{\small arctanh}\,b_{++}  }\, {\cal P}_+- {\cal P}_-,
\end{eqnarray}
which suggests that $ {\cal M}$ in the spinor representation would allow for
the analysis of BPS configurations in non-trivial Kalb-Ramond backgrounds along
the lines followed in \cite{Balasubramanian:1996uc} and \cite{Bergshoeff:1997kr}.

Further restrictions are found when requiring more supersymmetry.
Indeed, the existence of a second supersymmetry demands the
presence of two complex structures $J$ and $\bar J$ both
covariantly constant and such that the metric is hermitian with
respect to both of them. Furthermore, one of the two should be
expressed in terms of the other one, the metric, the Kalb-Ramond
field and the almost product structure. Just as for the case
without boundaries, no general manifest $N=2$ supersymmetric
description, involving only linear superfield constraints, 
can be given. However, we showed that at least the
type B K\"ahler models can be adequately described in $N=2$
superspace. It would be quite interesting to further investigate the $N=2$
superspace geometry, in particular for the A-type boundary conditions in the
presence of non-trivial $U(1)$ backgrounds.

An important application of the present paper would be the following. A crucial ingredient in the study of D-brane dynamics is its effective
action. While quite a lot is known about the effective action in the abelian case, at least in the limit of constant field strengths, only partial
results are known for the non-abelian case. In \cite{KS} (see also \cite{SW}), the effective action was obtained through fourth order in $ \alpha '$. This
analysis showed that derivative terms play an essential role in the non-abelian effective action and cannot be neglected. The complexity
of the results in \cite{KS} makes one wonder whether a closed expression to all order in $ \alpha '$ might ever be obtainable.
If such a closed expression exists, one should by taking the abelian limit, obtain a closed expression for the abelian effective
action {\em including all derivative corrections}! So before tackling the full non-abelian problem, it looks more reasonable to first study
the abelian case. Partial results, using various methods, were obtained in \cite{derivatives}. 
A powerful way to obtain the effective action passes over the calculation of the $\beta$-functions \cite{BI}. While all
calculations till now have been done in $x$-space, the results of the preceding section would allow a superspace calculation. The starting point would
be the action,
\begin{eqnarray}
\int d^2 \sigma d^2 \theta\, \sum_\alpha\left(-2i D Z^ \alpha \bar D Z^ {\bar \alpha }-8i
\Psi^ \alpha \Psi^{\bar \alpha }\right)+
\int d \tau d^2 \theta\, V(Z,\bar Z),\label{bgf}
\end{eqnarray}
with the chiral superfields defined in eq.\ (\ref{chiralsf}).
Taking Neumann boundary conditions in all directions and making a
background field expansion, would allow for a systematic analysis
of the $\beta$-functions directly in superspace. The simplicity of
eq.\ (\ref{bgf}) indicates that a systematic study of the
derivative correction to the effective action might be possible.

\bigskip

\acknowledgments

We thank Joanna Erdmenger, and in particular Cecilia Albertsson and 
Chris Hull for very useful and stimulating discussions.
This work was supported in part by the ``FWO-Vlaanderen''
through project G.0034.02, by the Federal Office for Scientific, Technical and
Cultural Affairs through the Interuniversity Attraction Pole P5/27 and by the
European Commission RTN programme HPRN-CT-2000-00131, in which the authors are
associated to the University of Leuven. When finishing this paper, a preprint appeared,
\cite{today}, where related issues, albeit using very different methods, are discussed.

\appendix

\section{Some conventions and notations}
We denote the worldsheet coordinates by $ \tau $ and $ \sigma $ and the light-cone
coordinates are,
\begin{eqnarray}
\sigma^{\pp} =  \tau +\sigma , \quad \sigma ^== \tau - \sigma \qquad \Rightarrow \qquad
\partial_\pp=\frac 1 2 ( \partial_ \tau + \partial_ \sigma ),\quad
\partial_==\frac 1 2 ( \partial_ \tau - \partial_ \sigma ).
\end{eqnarray}
The target-space coordinates are denoted by $X^a$, $a\in\{1,\cdots D\}$.
The target space data is encoded in the metric $g_{ab}(X)=g_{ba}(X)$ and the Kalb-Ramond
field $b_{ab}(X)=-b_{ba}(X)$. The torsion is given by the curl of the Kalb-Ramond field,
\begin{eqnarray}
T_{abc}=-\frac 3 2 b_{[ab,c]}.
\end{eqnarray}
We introduce two connections,
\begin{eqnarray}
\Gamma^{\ a}_{(\pm) bc}\equiv \big\{{}^a{}_{bc}\big\}\pm T^a{}_{bc},
\end{eqnarray}
where the first term is the standard Christoffel connection. The
connections are used to define covariant derivatives,
\begin{eqnarray}
\nabla^{(\pm)}_aV^b&=& \partial_a V^b+\Gamma^{\ b}_{(\pm) ca}V^c,\nonumber\\
\nabla^{(\pm)}_aV_b&=& \partial_a V_b-\Gamma^{\ c}_{(\pm) ba}V_c\,.
\end{eqnarray}
The curvature tensors are defined as,
\begin{eqnarray}
{[} \nabla_a^{(\pm)},\nabla_b^{(\pm)}{]}V^c=\frac 1 2 V^dR_{(\pm)dab}^{\ c}\pm
T^d{}_{ab}\nabla^{(\pm)}_dV^c,\label{int}
\end{eqnarray}
and we get explicitly,
\begin{eqnarray}
R^{\ a}_{(\pm)bcd}&=& \Gamma^{\ a}_{(\pm)bd,c}+ \Gamma ^{\ a}_{(\pm)ec}\Gamma ^{\ e}_{(\pm)bd}
-\ c\leftrightarrow d, \nonumber\\
R^{(\pm)}_{abcd}&=& \Gamma ^{(\pm)}_{abd,c}+\Gamma ^{(\pm)}_{ead} \Gamma ^{\ e}_{(\pm)bc}-\ c\leftrightarrow d.
\end{eqnarray}
The curvature tensors $R^{(\pm)}_{abcd}$ are anti-symmetric in the first and the last two indices, and they also
satisfy,
\begin{eqnarray}
R^+_{abcd}=R^-_{cdab}.
\end{eqnarray}

\section{The $N=(1,1)$ non-linear $ \sigma $-model}
Besides the target-space coordinates $X^a$ which are worldsheet scalars, we also have real worldsheet
fermions $\psi^a_+$ and $\psi^a_-$, which are target-space vectors. Including the auxiliary fields $F^a$,
the $d=2$, $N=(1,1)$ supersymmetry transformations are given by,
\begin{eqnarray}
&&\delta X^a = i \varepsilon ^+ \psi ^a_++i \varepsilon ^- \psi ^a_-, \nonumber\\
&& \delta  \psi ^a_+= - \varepsilon ^+ \partial_\pp \,X^a- \varepsilon ^- F^a, \nonumber\\
&& \delta \psi ^a_- = - \varepsilon ^- \partial_= X^a+ \varepsilon ^+F^a, \nonumber\\
&& \delta F^a= -i \varepsilon ^+ \partial_\pp \,\psi ^a_-
+i \varepsilon ^- \partial_= \psi ^a_+.\label{susy11}
\end{eqnarray}
The $d=2$ non-linear $ \sigma $-model is lagrangian is given by,
\begin{eqnarray}
{\cal L}&=&2(g_{ab}+b_{ab}) \partial_\pp \,X^a \partial_= X^b+
2i\, g_{ab}\,\psi ^a_+\nabla_=^{(+)} \psi ^b_+
+2i\, g_{ab}\,\psi ^a_-\nabla_\pp^{(-)} \psi ^b_- \nonumber\\
&&+R^{(-)}_{abcd} \psi _-^a \psi _-^b \psi _+^c \psi _+^d
+2 (F^a-i \Gamma ^{\ a}_{(-)cd} \psi ^c_- \psi ^d_+ )g_{ab}
(F^b-i \Gamma ^{\ b}_{(-)ef} \psi ^e_- \psi ^f_+ ).\label{lag11}
\end{eqnarray}
When there are no boundaries, one can show that the lagrangian eq.\ (\ref{lag11})
is invariant under the supersymmetry transformations
eq.\ (\ref{susy11}) without any further conditions.

\section{``Deriving'' the boundary superspace $N=1$ action from the bulk action}

The well-known superspace $N=(1,1)$ bulk action reads,
\begin{equation}
{\cal S}_{\rm bulk}= 2\, \int d \tau d\sigma D_+ D_- \left[(g_{ab}+b_{ab}) D_+ \Phi^a D_- \Phi^b\right] \, ,
\label{bulksup}
\end{equation}
where $D_+^2 = -i \partial_{\neq}$, $D_-^2= - i \partial_{=}$ and $\{D_+,D_-\}=0$.
Working out the superspace derivatives, one finds (\ref{lag11}).
This action is however {\em not} supersymmetric on the boundary.
We change coordinates to
\begin{eqnarray}
\theta & = & \theta^+ + \theta^-, \qquad D = \frac{1}{2} (D_{+}+D_{-}) \, , \nonumber \\
\tilde{\theta} & = & \theta^+ - \theta^-, \qquad \tilde{D} = \frac{1}{2}(D_{+}-D_{-}) \, ,
\end{eqnarray}
where $D^2=\tilde{D}^2= -\frac{i}{4}\frac{\partial}{\partial \tau}$ and, importantly
$\{D,\tilde{D}\}= - \frac{i}{2}\frac{\partial}{\partial \sigma}$.
Consider the following ``improved'' action:
\begin{eqnarray}
{\cal S} & = & -4 \int d \tau d \sigma D \tilde{D} \left[(g_{ab}+b_{ab}) D_+ \Phi^a D_- \Phi^b\right] \nonumber \\
         & = & -4 \int d \tau d \sigma d \theta \tilde{D} \left[(g_{ab}+b_{ab}) D_+ \Phi^a D_- \Phi^b\right] \, .
\label{improvedbulksup}
\end{eqnarray}
This is explicitly supersymmetric on the boundary and, using $D_+
D_-= -2D \tilde{D} - \frac{i}{2} \frac{\partial}{d\sigma}$, modulo
boundary terms equivalent to (\ref{bulksup}). Moreover,
identifying
\begin{eqnarray}
\Phi^a & = & X^a \, , \nonumber \\
\tilde{D} \Phi^a & = & - i \Psi^a \, ,
\label{bulktoboundary}
\end{eqnarray}
and working out the $\tilde{D}$ derivative, one finds
the model (\ref{Sd1}),(\ref{res1}),(\ref{res2}) and (\ref{res3}).

Another possibility is the action (\ref{improvedbulksup}) with $D$ and $\tilde{D}$ interchanged.
Using (\ref{bulktoboundary}) in (\ref{id1}),
\begin{eqnarray}
\psi ^a_+&=& i (D\Phi^a + \tilde{D}\Phi^a), \nonumber\\
\psi^a_-&=&i \eta (D\Phi^a -\tilde{D}\Phi^a) \, ,
\end{eqnarray}
one sees that this is equivalent to putting $\eta \rightarrow -\eta$.

Of course, the argument in the preceding paragraphs does not replace the exhaustive analysis of section \ref{N1superspace},
because it is not a priori clear that the most general boundary model could be written in the form (\ref{improvedbulksup}). In fact, this
procedure does not work when one tries to derive $\sigma$-models in $N=2$ boundary
superspace from the $N=(2,2)$ superspace $\sigma$-models.

We proceed to derive the second supersymmetry from the form of this symmetry in the bulk,
\begin{equation}
\delta \Phi^a=\hat{\varepsilon}_+ J^a{}_b D_{+} \Phi^b + \hat{\varepsilon}_- \bar{J}^a{}_b D_{-} \Phi^b \, .
\label{bulk2ndsusy}
\end{equation}
Since only one of these symmetries will survive on the boundary we put
$\hat{\varepsilon}_+=\hat{\varepsilon}_-=\frac{1}{2}\hat\varepsilon$.  Using (\ref{bulktoboundary})
we find for the bottom and top components of (\ref{bulk2ndsusy}),
\begin{eqnarray}
\delta X^a &=&\hat\varepsilon {\cal J}^{\ a}_{(1)b}(X)DX^b+ \hat\varepsilon{\cal J}^{\ a}_{(2)b}(X) \Psi  ^b, \nonumber\\
\delta \Psi ^a &=& \hat\varepsilon{\cal K}^{\ a}_{(1)b}(X)D \Psi ^b+ \hat\varepsilon{\cal K}^{\ a}_{(2)b}(X) \dot X ^b+
\hat\varepsilon{\cal K}^{\ a}_{(3)b}(X)X'{}^b \nonumber \\
&& \hat\varepsilon{\cal L}_{(1)bc}^{\ a}(X) \Psi ^b \Psi ^c + \hat\varepsilon{\cal L}_{(2)bc}^{\ a}(X) \Psi ^b DX^c
+ \hat\varepsilon{\cal L}_{(3)bc}^{\ a}(X) DX ^b DX^c \, , \label{2ndsusyfull}
\end{eqnarray}
but now with,
\begin{eqnarray}
{\cal J}_{(1)} & = & \frac{1}{2}(J +\bar{J}), \quad i {\cal J}_{(2)}=-4 {\cal K}_{(2)}= \frac{1}{2}(J-\bar{J}), \nonumber \\
{\cal K}_{(1)} & = & \frac{1}{2}(J +\bar{J}), \quad  {\cal K}_{(3)} = -\frac{1}{4} (J+\bar{J}), \nonumber \\
{\cal L}^{a}_{(1)bc}&=&  \frac{i}{2} \left(\partial_{[b} J^a{}_{c]}-\partial_{[b} \bar{J}^a{}_{c]}\right),  \nonumber \\
{\cal L}^{a}_{(2)bc}&=&- \frac{1}{2} \left(\partial_{b} J^a{}_{c}+\partial_{b} \bar{J}^a{}_{c} \right), \nonumber\\
{\cal L}^{a}_{(3)bc} &=& 0, \label{2ndsusyfull2}
\end{eqnarray}
while $J$ and $\bar{J}$ are the same tensors as in section \ref{N2susy}.
There is however more freedom than in the bulk. In section \ref{N2susy} the most general susy transformation leaving
the action invariant and for which the algebra closes off-shell, was considered.
When requiring only on-shell closure one finds that one can add the following transformation to (\ref{2ndsusyfull}),
\begin{equation}
\delta \Psi ^a = \hat{\varepsilon} K^a{}_b \left( \frac{1}{2} D \Psi^b - \frac{1}{8} X^b{}'
+ \frac{1}{2} K^{da} \{dbc\} \Psi^b DX^c + \frac{i}{4} K^{da} T_{dbc} \left(\Psi^b \Psi^c + DX^b DX^c \right) \right) \, ,
\label{eomterm}
\end{equation}
where $K$ is an arbitrary antisymmetric tensor, $g_{a(b}K^a{}_{c)}=0$.
Note that $K$ multiplies the dimension-$1$ equation of motion, so that this transformation vanishes on-shell.
So the most general susy transformation looks like,
\begin{eqnarray}
\delta X^a &=&\hat{\varepsilon} {\cal J}^{\ a}_{(1)b}(X)DX^b+ \hat{\varepsilon}{\cal J}^{\ a}_{(2)b}(X) \Psi  ^b, \nonumber\\
\delta \Psi ^a &=& \hat{\varepsilon}{\cal K}^{\ a}_{(1)b}(X)D \Psi ^b+ \hat{\varepsilon}{\cal K}^{\ a}_{(2)b}(X) \dot X ^b
+ \hat{\varepsilon}{\cal K}^{\ a}_{(3)b}(X) X^{'b} \nonumber \\
& + & \hat{\varepsilon}{\cal L}_{(1)bc}^{\ a}(X) \Psi ^b \Psi ^c + \hat{\varepsilon}{\cal L}_{(2)bc}^{\ a}(X) \Psi ^b DX^c
+ \hat{\varepsilon}{\cal L}_{(3)bc}^{\ a}(X) DX ^b DX^c ,\label{2ndsusya}
\end{eqnarray}
with,
\begin{eqnarray}
{\cal J}_{(1)} & = & \frac{1}{2}(J +\bar{J}), \quad i {\cal J}_{(2)}=-4 {\cal K}_{(2)}= \frac{1}{2}(J-\bar{J}) \, ,
\nonumber \\
{\cal K}_{(1)}&=&\frac{1}{2}(J+\bar{J}+K), \quad {\cal K}_{(3)}= -\frac{1}{8}(2J+2\bar{J}+K) \, , \nonumber \\
{\cal L}^{a}_{(1)bc}&=& - \frac{1}{2} \left(\partial_{[b} J^a{}_{c]}+\partial_{[b} \bar{J}^a{}_{c]}\right)
+ \frac{i}{4} g^{ad}\; K^e{}_d \; T_{bce} \, ,  \nonumber \\
{\cal L}^{a}_{(2)bc}&=& \frac{i}{2} \left(\partial_{b} J^a{}_{c}+\partial_{b} \bar{J}^a{}_{c}-i g^{ad}\;K^e{}_d \;\big\{ebc\big\} \right) \, ,
\nonumber\\
{\cal L}^{a}_{(3)bc} &=& \frac{i}{4} g^{ad}\;K^e{}_d \; T_{bce} \, .\label{2ndsusyb}
\end{eqnarray}
The algebra closes off-shell if and only if,
\begin{itemize}
\item ($[J,\bar{J}]=0$ and $K=0$) or
\item $K=-2J -2 \bar{J}$.
\end{itemize}
The latter leads to eq.\ (\ref{2ndsusy}) with eq.\ (\ref{2ndsusy2}).

\end{document}